\documentclass[amsfonts, amssymb, amsmath, showkeys, nofootinbib, twocolumn,superscriptaddress]{revtex4-1}

\usepackage{amsthm}
\usepackage[english]{babel}
\usepackage[utf8]{inputenc}

\usepackage{mathtools}
\usepackage{physics}
\usepackage{xcolor}
\usepackage{graphicx}
\usepackage{adjustbox}
\usepackage{placeins}
\usepackage[T1]{fontenc}
\usepackage{lipsum}
\usepackage{csquotes}
\usepackage{bm}
\usepackage{float}
\usepackage{multirow}

\usepackage[pdftex, pdftitle={Article}, pdfauthor={Author}]{hyperref} 
\begin{document}

\title{Quasi-two-dimensional ferromagnetism and anisotropic interlayer couplings in the magnetic topological insulator MnBi$_{2}$Te$_{4}$}

\author{Bing Li}
\affiliation{Ames Laboratory, Ames, IA, 50011, USA}
\affiliation{Department of Physics and Astronomy, Iowa State University, Ames, IA, 50011, USA}

\author{D. M. Pajerowski}
\affiliation{Oak Ridge National Laboratory, Oak Ridge, TN, 37831, USA}

\author{S. X. M. Riberolles}
\affiliation{Ames Laboratory, Ames, IA, 50011, USA}

\author{Liqin Ke}
\affiliation{Ames Laboratory, Ames, IA, 50011, USA}

\author{J.-Q.~Yan}
\affiliation{Oak Ridge National Laboratory, Oak Ridge, TN, 37831, USA}

\author{R.~J.~McQueeney}
\affiliation{Ames Laboratory, Ames, IA, 50011, USA}
\affiliation{Department of Physics and Astronomy, Iowa State University, Ames, IA, 50011, USA}

\date{\today} 

\begin{abstract}

MnBi$_2$Te$_4$ (MBT) is a promising van der Waals layered antiferromagnetic (AF) topological
insulator that combines a topologically non-trivial inverted Bi-Te band gap with ferromagnetic (FM)
layers of Mn ions. Inelastic neutron scattering on single crystals reported here describes rather complex magnetism in MBT.
The magnetic anisotropy that controls the bulk and surface magnetic field response of MBT is found to have contributions from both single-ion and interlayer two-ion terms.
A description of the quasi-two-dimensional intralayer FM spin waves requires long-range, competing FM and AF interactions and anomalous damping.  While this might suggest carrier mediated magnetic coupling, $\textit{ab initio}$ calculations in insulating MBT also find long-range interactions and classical spin dynamics simulations suggest that magnetic vacancies are at least partially responsible for observations of anomalous damping near the zone boundary.

\end{abstract}


\maketitle

Intrinsic magnetic topological materials are of interest for potential applications in dissipationless quantum transport, optical, and magnetoelectric responses. These materials couple topological electronic bands with magnetism such that the symmetry of the magnetic state determines the topological phase.  Ideally, the magnetic symmetry can be easily manipulated with external fields or chemical doping, resulting in access to, or switching between, unique topological phases.

One useful structural motif for these purposes consists of alternating magnetic and topological layers, as found in the MnBi$_{2}$Te$_{4}$ (MBT) family of materials \cite{Eremeev17, Otrokov18, Otrokov18_2, Zhang19, Lee19, Yan19}.  These materials are closely related to Bi$_2$Te$_3$ topological insulators and a single septuple block of MBT contains a ferromagnetic (FM) Mn layer sandwiched by Bi-Te layers that host inverted topological electronic bands. Without magnetic order, MBT is a strong topological insulator with gapless Dirac surface states protected by time-reversal symmetry ($\Theta$). When $\Theta$ is broken by magnetic ordering, the magnetic coupling between FM Mn layers controls the emergent quantum anomalous Hall (FM coupling) or axion insulator (AF coupling) topological states. In the AF axion insulator case, the preserved symmetry $S=\Theta t_{1/2}$, where $t_{1/2}$ is the translation vector between oppositely magnetized Mn layers, admits $Z_2$ topological classification \cite{Mong10}.

The van der Waals nature of the bonding between septuple blocks in MBT causes weak magnetic couplings between Mn layers. In bulk MBT, this AF coupling results in an A-type AF ground state (FM triangular Mn layers that stack AF) with $T_{\rm N}=24$ K \cite{Yan19, Ding20}, however, switching from A-type AF to a fully-polarized FM state is easily achievable in relatively weak applied fields \cite{Otrokov18_2, Yan19, Tan20}.  Also, the relative ease of exfoliating septuple layers allows for studies of monolayer and few-layer devices \cite{Otrokov19, Gong19, Liu20}.  This insight has lead to the demonstration of the quantum anomalous Hall effect in odd-layer devices with an uncompensated net magnetization \cite{Deng20}.

Despite many recent successes that demonstrate the promise of the MBT-family of materials, there remain some mysteries and open questions.  In the A-type AF, Dirac cones are expected to be gapped on (001) surfaces, where $S$ is broken. However, ARPES results have not clearly observed the gapped (001) surface state, possibly due to magnetic disorder or changes in the magnetic structure near the surface \cite{Chen19, Hao19, Swatek19}.  This suggests that the magnetic state of the Mn surface layer may not be well understood.  Inelastic powder neutron scattering data on MBT found competing FM and AF intralayer interactions that provide a source of magnetic frustration in the bulk material \cite{Li20}.  Also, there are reports of bulk $n$-type charge carriers in MBT \cite{Yan19, Lee19}, presumably resulting from intrinsic defects, whose transport properties are strongly influenced by magnetism. Conduction electrons can mediate exchange paths between local Mn moments that are not present in the insulator. Furthermore, chemical impurities and disorder, in the form of Mn-Bi antisite mixing and magnetic vacancies, result in magnetic dilution and ferrimagnetism within a single septuple block, as recently observed in MnSb$_2$Te$_4$ \cite{Murakami19}.

In this Letter, we use inelastic neutron scattering (INS) and density-functional theory (DFT) to study the magnetic interactions in MBT in great detail. Heisenberg model analysis of the INS data finds that intralayer interactions are the dominant energy scale, consistent with a quasi-two-dimensional van der Waals (vdW) magnet. The AF interlayer coupling and uniaxial single-ion anisotropy are of comparable strength, and the introduction of an anisotropic interlayer coupling term provides more consistent modeling of both the spin waves and the field-dependent metamagnetic transitions \cite{Otrokov18, Yan19, Tan20}. The intralayer dispersion of magnons above the spin gap is surprisingly almost linear at the Brillouin zone center and throughout the zone, rather than the expected quadratic dispersion. This necessitates the inclusion of long-range intralayer interactions and confirms previous reports of competing FM and AF intralayer interactions \cite{Li20} that are supported by DFT calculations for insulating MBT. We also find anomalous broadening of the intralayer spin waves, which qualitatively resembles the calculated magnon life-time due to electron-magnon coupling in FM semiconductors \cite{Woolsey1970}. However, classical spin dynamics simulations indicate that magnetic vacancies caused by antisite mixing also contribute to the spectral broadening. Overall, our results highlight the complexity of magnetism in MBT, which includes long-range interactions, multiple contributions to the magnetic anisotropy, and strong spectral broadening with possible origins in magnetic defects and coupling to charge carriers.

\begin{figure}
    \includegraphics[width=0.95\linewidth]{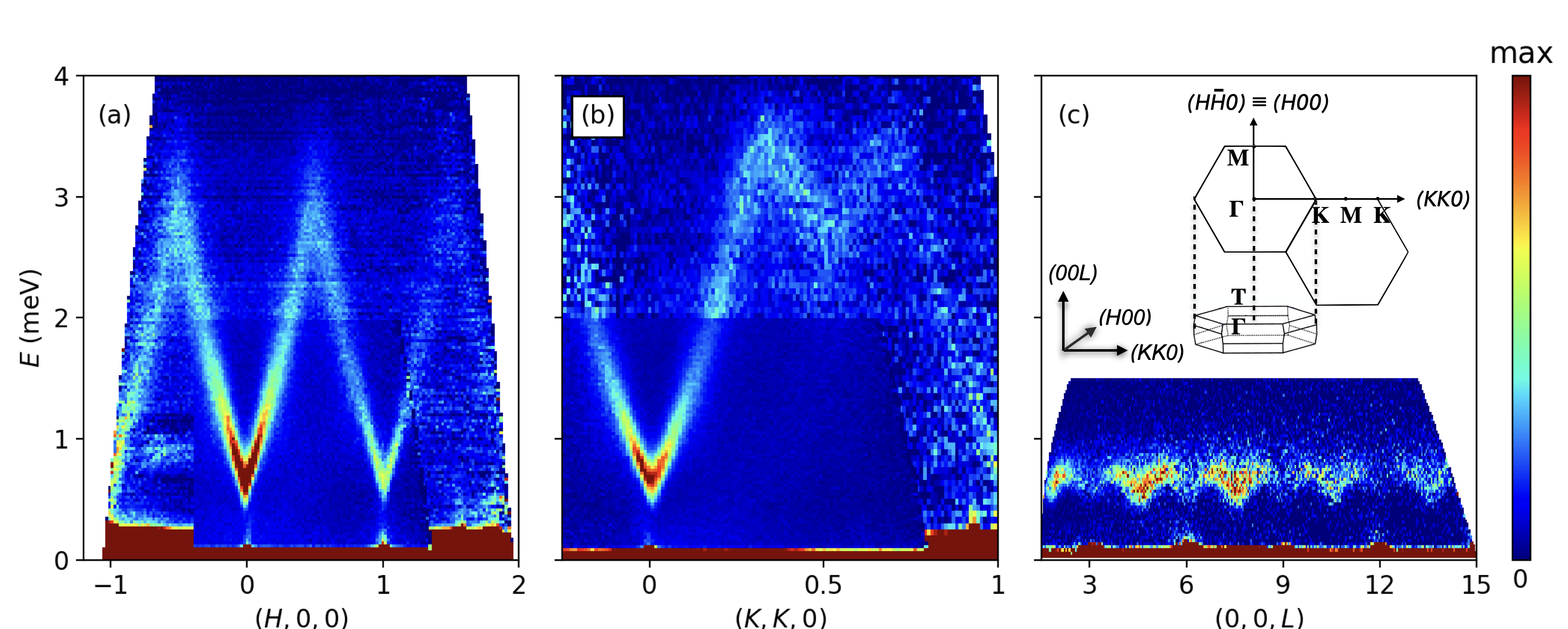}
    \caption{\footnotesize Representative raw data plots of spin waves at $T=$ 1.8 K, composed of high-resolution $E_\mathrm{i}=$ 3.3 meV data superposed on top of the coarser-resolution $E_\mathrm{i}=$ 6.6 meV data, along the hexagonal (a) $(H,0,0)$, (b) $(K,K,0)$, and (c) $(0,0,L)$ directions. Square brackets are used to denote the range of summation in $\bm Q$. In (a) and (b), $L=[0, 20]$ for 6.6 meV data and $[3,12]$ for 3.3 meV data. In (a), $K=[-0.05, 0.05]$, In (b) $H=[-0.05, 0.05]$, In (c), $K=[-0.02,0.02]$, $H=[-0.02,0.02]$. The inset to panel (c) shows the rhombohedral Brillouin zone and its relation to the 2D hexagonal zone.}
    \label{fig:fig1}
\end{figure}

Single-crystal samples of MnBi$_2$Te$_4$ were grown out of Bi$_2$Te$_3$ flux \cite{Yan19}. The crystals are slightly Bi-rich with a composition of Mn$_{0.96(2)}$Bi$_{2.04(2)}$Te$_{4.00(2)}$.  MBT crystallizes in space group R-3m (\#166), with lattice parameters $a=b=4.33$~\AA, $c=40.91$~\AA. SQUID measurements confirm that AF ordering occurs at $T_{\text{N}}=24.7$~K and neutron diffraction finds a magnetic propagation vector of $\bm{\tau}=(0,0,3\pi/c)$ consistent with the A-type order. Throughout the manuscript, we use hexagonal notation to describe the reciprocal space vectors $\bm{Q}=\frac{2\pi}{a}(H\hat{a}+K\hat{b})+\frac{2\pi}{c}L\hat{c}$ and special points [$\Gamma=(0,0,0)$, $M=(1/2,0,0)$, $K=(1/3,1/3,0)$, and $T=(0,0,3/2)$] rather than rhombohedral notation.  The rhombohedral Brillouin zone and its relation to the 2D hexagonal zone is shown in the inset to Fig.~\ref{fig:fig1}(c).

The INS measurements were performed at the Cold Neutron Chopper Spectrometer (CNCS) at the Spallation Neutron Source at the Oak Ridge National Laboratory.  Separate measurements were performed with the sample mounted to a helium cryostat and oriented with either a $(H,0,L)$ or $(K,K,L)$ horizontal scattering plane. For each orientation, measurements were performed using incident energies of $E_\mathrm{i}$= 3.3 meV and 6.6 meV while the sample was rotated about the vertical axis for $\bm{Q}$--space coverage. The sample was cooled to base temperature $T$ = 1.8 K for all measurements \cite{SM}.

Figures~1(a)--(c) show INS intensities of MBT [proportional to the dynamic spin-spin correlation function $S(\bm{Q},E)$] as a function of energy along $(H,0,0)$, $(K,K,0)$ and $(0,0,L)$ directions. The data reveal a spin gap with energy $\Delta=0.569(15)$ meV (The onset of the gap $\Delta_T\approx$~0.4 meV, corresponds to the spin-flop field of 3.4 T\cite{Otrokov18_2, Yan19, Lee19}. See supplemental materials (SM) for details). Along $(0,0,L)$ Fig.~1(c) shows that the dispersion is shallow due to weak interlayer interactions across the vdW gap. Intensity maxima along $(0,0,L)$ occur at $L=3n \pm \frac{3}{2}$ due to the staggered magnetization of the A-type structure along $c$.  The in-plane dispersions conform, to some degree, to expectations for a FM triangular layer.  However, rather than a smooth cosine-like dispersion with quadratic behavior near the gap, Fig.~1(a) displays a sawtooth pattern along $(H,0,0)$ and maintains relatively linear dispersion across the Brillouin zone to the $M$--point.  In Fig.~1(b), a similar linear dispersion is also observed in the $\Gamma-K$ section along ($K,K,0$) which continues along the $K-M-K$ zone edge to form an "M"--shape. Within a Heisenberg model described below, linear dispersion can be accommodated by the inclusion of long-range interactions with many Fourier components contributing to the $\bm{Q}$-dependent exchange function, $J(\bm{Q})$.  In addition, visual inspection (more clear in Fig.~3) suggests significant spectral broadening of the intralayer spin wave modes, especially close to the Brillouin zone boundary.

\begin{figure}
    \includegraphics[width=0.99\linewidth]{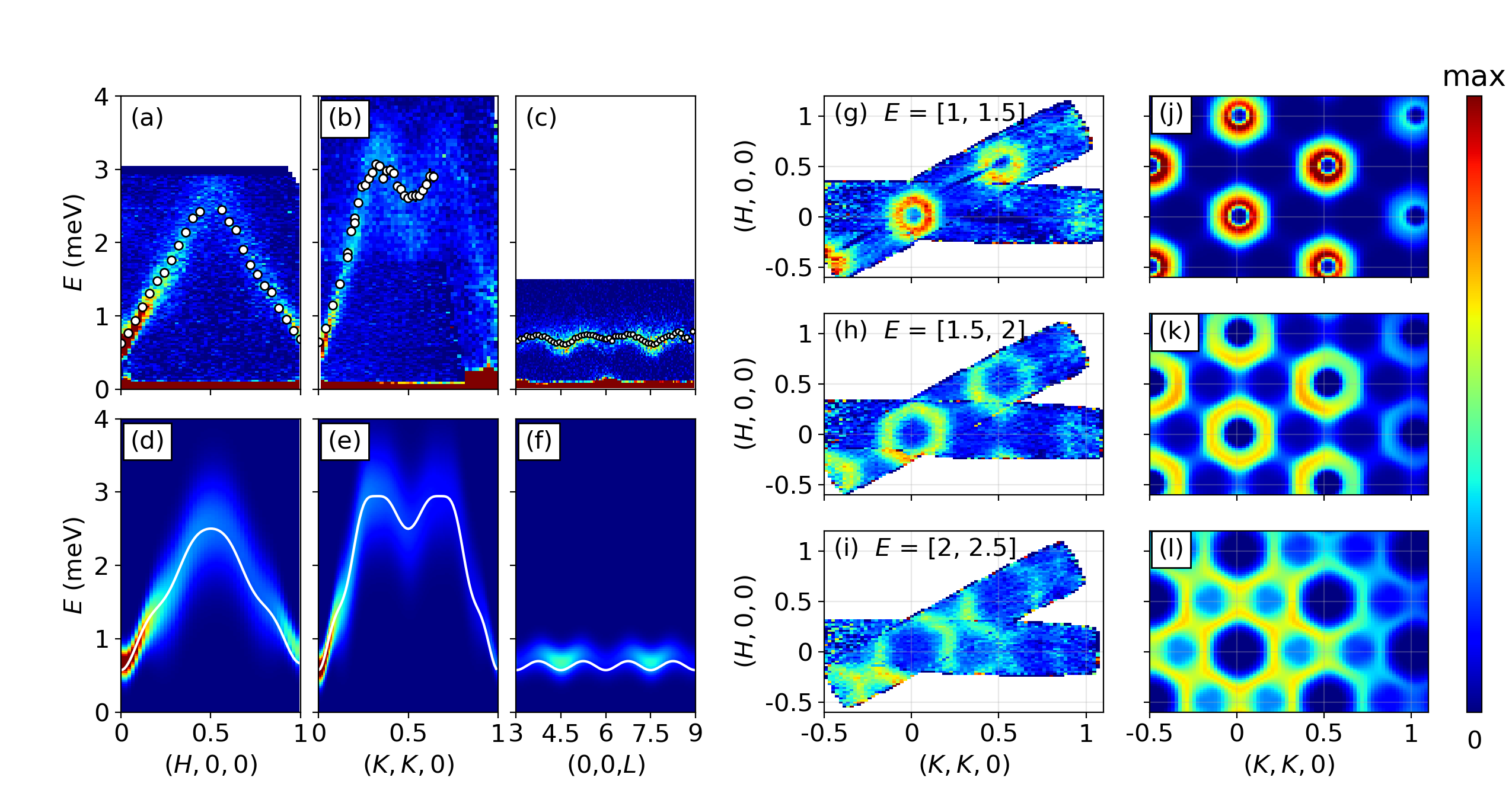}
    \caption{\footnotesize (a)--(c) Experimental neutron intensities along the $(H,0,0)$, $(K,K,0)$, and $(0,0,L)$ directions, respectively. To get accurate intralayer dispersions, narrow $L$ ranges centered at $\Gamma$ and T points are summed over for better statistics. White symbols correspond to the experimentally determined spin wave dispersion, $\hbar\omega(\bm{Q})$ obtained from fits to constant-$\bm{Q}$ energy cuts. (d)--(f) Heisenberg model calculations of the neutron intensities along the $(H,0,0)$, $(K,K,0)$, and $(0,0,L)$ directions with white lines indicating the model dispersion relation. (g)--(i) Constant-$E$ slices of the experimental data in the $(H,K,0)$ plane at energies of 1.25, 1.75 and 2.25 meV, respectively, summed from $L=$[-10, 30]. (j)--(l) Heisenberg model calculations of the same energy cuts as shown in panels (g)--(i).  Heisenberg model calculations use parameters described in the main text and are convoluted with the experimental linewidth, $\gamma(\bm{Q})$. See SM for details.}
    \label{fig:disp}
\end{figure}

To obtain direct information about the various magnetic energy scales and potential long-range interactions, one-dimensional cuts along $E$ at selected $\bm{Q}$ positions were fit to a Gaussian lineshape. Peak centers determine the experimental spin wave dispersion $\hbar\omega(\bm{Q})$.
The full widths at half maximum (FWHM) $\gamma(\bm{Q})$ show a  non-trivial $\bm{Q}, E$-dependence, and are parameterized by an analytical function of arctan($E$) that is used for subsequent simulation of the spin wave intensities\cite{SM}.

The extracted values of $\hbar\omega(\bm{Q})$, which are shown in Fig.~2(a)--(c), are fit to linear spin wave theory (LSWT) based on the Heisenberg model
\begin{equation}
    \mathit{H} = -\sum_{\langle ij \rangle,\parallel} J_{ij} \bm{S}_{i} \cdot \bm{S}_{j} -  J_{\text{c}} \sum_{\langle ij \rangle,\perp}\bm{S}_{i} \cdot \bm{S}_{j} - D \sum_{i}(S^z_i)^2 +  \mathit{H_c^{\text{aniso}}},
\label{heisenberg}
\end{equation}
where $i$ labels the Mn ion at position $\bm{R}_i$ with spin $\bm{S}_i$, $J_c$ is the nearest-neighbor interlayer exchange (See SM for the discussion about effective $J_{\text{c}}$ from multiple interlayer couplings), $J_{ij}$ are the pairwise intralayer exchanges, $D$ is the uniaxial single-ion anisotropy, and $\mathit{H_{\text{c}}^{\text{aniso}}}=-J_{{\text{c}}}^{\text{aniso}} \sum_{\langle ij \rangle,\perp}{S}_{i}^{z}{S}_{j}^{z}$ accounts for the anisotropic contribution in the interlayer interaction. We use reduced--$\chi^2$ analysis to compare the experimental and calculated LSWT dispersions given by $\hbar\omega(\bm{Q}) = \sqrt{A(\bm{Q})^2 - B(\bm{Q})^2}$ where
\begin{align*}
A(\bm{Q}) &= S(J(\bm{Q})+\frac{1}{2}J(\bm{Q}+\bm{\tau})+\frac{1}{2}J(\bm{Q}-\bm{\tau})-2J(\bm{\tau}))\nonumber \\
&+2SD+ZSJ_{{\text{c}}}^{\text{aniso}},\nonumber\\
B(\bm{Q}) &= S(J(\bm{Q})-\frac{1}{2}J(\bm{Q}+\bm{\tau})-\frac{1}{2}J(\bm{Q}-\bm{\tau}))\label{AqBq}.
\end{align*}
and $J(\bm{Q}) = \sum_{j} J_{0j} e^{i \bm{Q} \cdot \bm{R}_j}$ is the Fourier transformation of all neighboring magnetic interactions for a given atom, $\bm{\tau} = (0,0,\frac{3}{2})$ is the magnetic propagation vector, and $Z=6$ is the interlayer nearest-neighbor coordination number. The LSWT dynamical structure factor is proportional to the transverse components of the dynamic spin-spin correlation function, $S^{xx}(\bm{Q}, E)= S^{yy}(\bm{Q}, E)=S(A(\bm{Q})-B(\bm{Q}))/\hbar\omega(\bm{Q})$, and

\begin{equation}
   S(\bm{Q},E)=S |f(\bm{Q})|^{2}\frac{A(\bm{Q})-B(\bm{Q})}{\hbar\omega(\bm{Q})}(1+\hat Q_z^2) \delta(E-\hbar\omega(\bm{Q}))
   \label{Sqw}
\end{equation}
where $f(\bm{Q})$ is the magnetic form factor of Mn$^{2+}$ ion and $\hat Q_z$ is the z component of the unit vector in $\bm Q$ direction.

To compensate for the effect of finite $\bm{Q}$--sampling, an identical analysis of one-dimensional cuts along $E$ were made for the model calculation. The values of magnetic interactions are determined when the Gaussian peak centers in the data and simulation best agree with each other. Figure~\ref{fig:disp} compares the experimental neutron intensities and LSWT simulations along the principal reciprocal space directions [Figs.~2(a)--(f)] as well as constant energy contours at selected energy transfers within the plane[Figs.~2(g)--(l)]. Overall, the Heisenberg model provides a good agreement between the data and the simulation in all momentum-space directions.

As shown in Figs.~2(c) and 2(f) for the dispersion along $(0,0,L)$, an isotropic nearest-neighbor interlayer AF exchange $SJ_{\text{c}}=-0.065(2)$~meV and single-ion anisotropy $SD=0.150(5)$~meV provide reasonable agreement between the model and INS data. The parameters are consistent with previous results obtained from powder INS \cite{Li20}. However, this set of parameters would predict incorrect saturation fields $H_{\text{sat}}^{\text{c}}$ and $H_{\text{sat}}^{\text{ab}}$, and the introduction of $J_{c}^{\text{aniso}}$ resolves this inconsistency. Note that $J_{c}^{\text{aniso}}$ does not affect the spin wave bandwidth along $(0,0,L)$, it only affects the gap. Modelling based on magnetization data gives $SD=0.070$~meV, $SJ_{c}=-0.075$~meV and $SJ_{c}^{\text{aniso}}=-0.026$~meV \cite{SM}. To confirm value of $SD$ obtained from magnetization data, we further conduct INS measurements on Bi$_{1.99}$Mn$_{0.01}$Te$_{3}$, where Mn ions share a similar local environment as MBT. The observed splittings of spin $S=5/2$ multiplets give a direct measurement of $D=0.0325(3)$ meV and $SD=0.0814(7)$ meV. By fixing this value of SD, fits to the INS data confirm the presence of an anisotropic interlayer exchange coupling with $SJ_{c}=-0.065(2)$~meV and $SJ_{c}^{\text{aniso}}=-0.023(3)$~meV. (See SM for experimental details and DFT estimation of $SD$ and  $SJ_{c}^{\text{aniso}}$.)

An optimal Heisenberg model description of the unusual intralayer dispersion requires the addition of long-range pairwise interactions. We fix $SJ_{\text{c}}$, $SJ_{\text{c}}^{\text{aniso}}$ and $SD$ and study the behavior of the reduced--$\chi^2$ by adding successive intralayer interactions. Fits improve noticeably up to the seventh nearest-neighbor and then deteriorate. The best-fit intralayer exchange values (in meV) up to the seventh neighbor are: $[SJ_1,~SJ_2,~SJ_3,~SJ_4,~SJ_5,~SJ_6,~SJ_7]$~= [0.233(2), $-0.033(2)$, 0.007(2), $-0.003(2)$, 0.016(2), 0.013(2), 0.008(1)]. The dependence of the reduced--$\chi^2$ with an increasing number of neighbor shells and other details on the fitting are included in SM.

\begin{figure}
    \includegraphics[width=\linewidth]{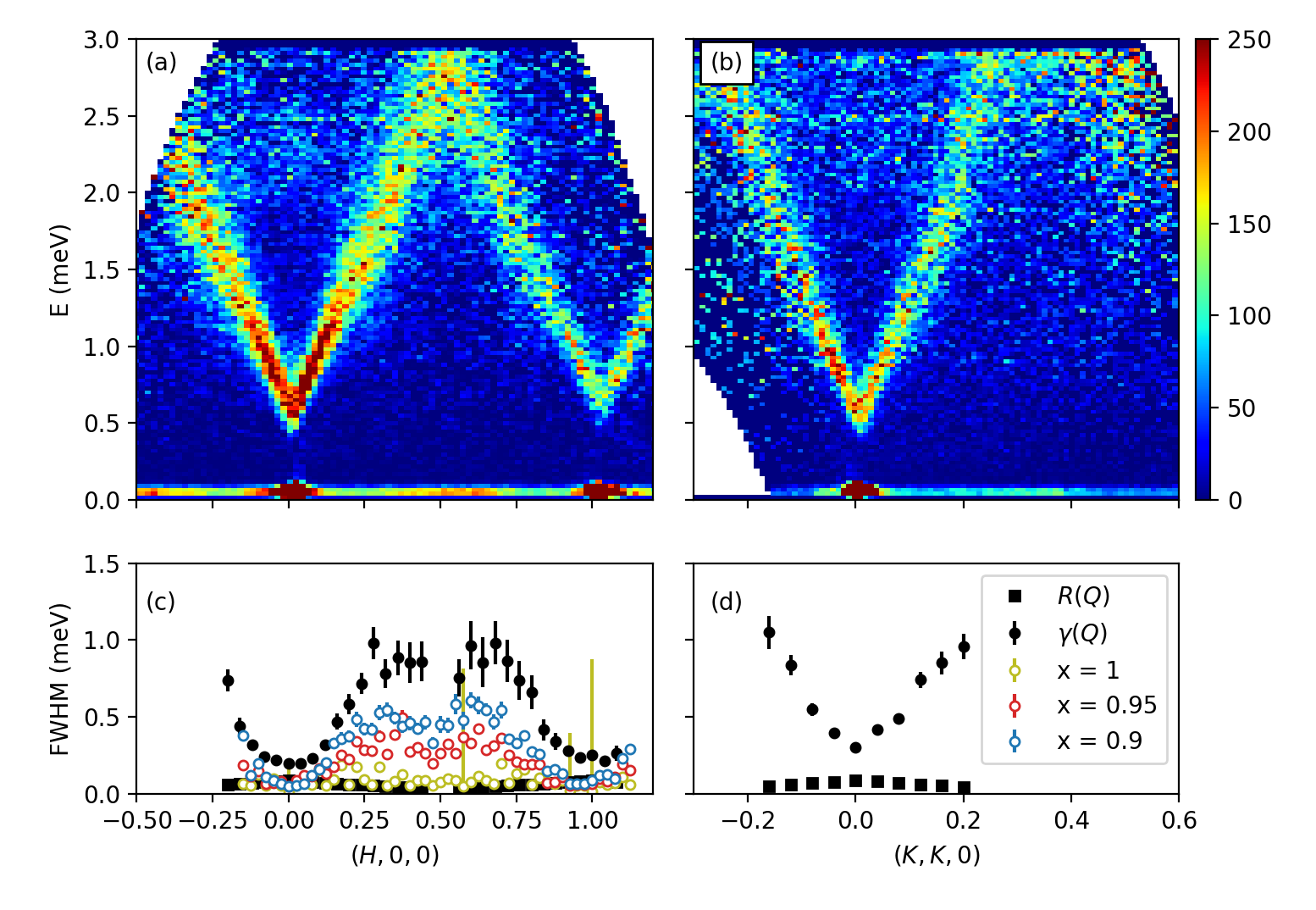}
    \caption{\footnotesize Experimental neutron intensities along (a) $(H,0,0)$ and (b)$(K,K,0)$, plotted as $E\cdot S(\bm{Q},E)/f^2(\bm{Q})$ to equalize the intensities at the top and bottom of the band. Fitted peak widths ($\gamma(\bm{Q})$, filled circles) and resolution-limited peak widths ($R(\bm{Q})$, filled squares) are displayed along (c) $(H,0,0)$ and (d) $(K,K,0)$. Open circles are from numerical simulations for the spin dynamics of Mn$_x$Bi$_2$Te$_4$ with different concentrations of magnetic vacancies (See SM \cite{SM}).}
    \label{fig:damping}
\end{figure}

The simulated intensities also allow for deeper analysis of the spin wave lifetimes. In Figs.~\ref{fig:damping}(a) and (b), the contour plots show scaled intensities $ES(\bm{Q},E)/|f(\bm{Q})|^2$ along $(H,0,0)$ and $(K,K,0)$ where scaling removes the trivial $E$ and $\bm{Q}$ dependencies to emphasize the intensity contrast. Figures~\ref{fig:damping}(c) and (d) compare the experimental Gaussian linewidths $\gamma(\bm{Q})$, with the linewidths $R(\bm{Q})$, obtained from LSWT simulated intensities assuming the intrinsic linewidths are resolution-limited. The contribution of finite $\bm{Q}$--sampling of the dispersion to $R(\bm{Q})$ is also taken into consideration. Here data with narrow $\bm{Q}$ ranges centered at $\Gamma$ and T points were summed together to improve statistics.

In principle, contributions to $\gamma(\bm{Q})$ that exceed $R(\bm{Q})$ come from intrinsic sources of broadening. The intrinsic broadening is substantial everywhere in the Brillouin zone, especially near the zone boundary where it reaches a maximum of $\gamma(\bm{Q})/\hbar\omega(\bm{Q}) \approx$~0.6 near $(0.3,0,0)$. Close to the zone center, the linewidths increase approximately quadratically, $\gamma(\bm{Q}) \propto Q^2$. Intrinsic line broadening can arise from several sources, including the coupling of spin waves to phonons or electrons, non-linear magnon-magnon coupling, or disorder. We find $\bm{Q}$-dependence of $\gamma(\bm{Q})$ qualitatively resembles the calculated magnon life-time due to electron-magnon coupling in FM semiconductors \cite{Woolsey1970}. However, we also find that magnetic vacancies contribute to the spectra broadening. Simulations of classical Landau-Lifshitz-Gilbert spin dynamics using the UppASD package \cite{Eriksson08} were used to extract the linewidths for different Mn vacancy concentrations, as plotted in Fig.~\ref{fig:damping}(c). We find that vacancy concentrations appropriate for our MBT sample ($\approx 10\%$) introduce significant contribution to the anomalous broadening (See SM). This suggests that the possibility to study potential electron-magnon coupling in MBT would require crystals with lower antisite mixing, or modified carrier concentration (e.g., by Sb substitution).

\begin{figure}[ht]
    \includegraphics[width=\linewidth]{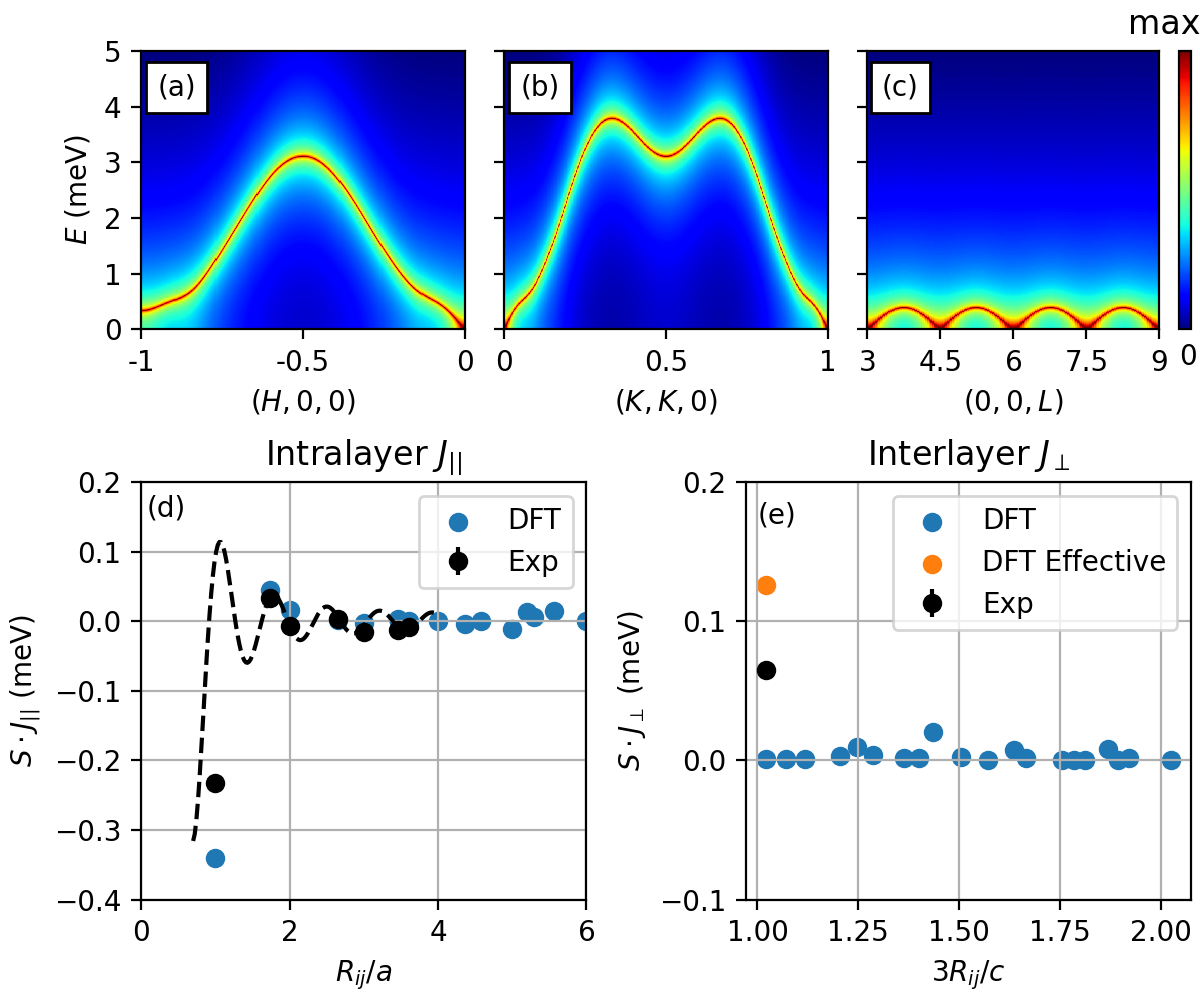}
    \caption{\footnotesize (a--c) The imaginary part of the calculated dynamic transverse spin susceptibility $\mathrm{Im}[\chi^{+-}(\bf{Q}),\it{E}]$ along (a) $(H,0,0)$, (b) $(K,K,0)$ and (c) $(0,0,L)$, the intensity is in logarithmic scale. (d--e) Calculated exchange parameters of (d) intralayer $J_{\parallel}$ and (e) interlayer $J_{\perp}$ as a function of distance in units of corresponding lattice parameters, compared to those extracted from INS experiment. Dashed line is a guide to the eye.}
    \label{fig:DFT}
\end{figure}

To understand of the origin of the magnetic interactions, we calculate the dynamic transverse spin susceptibility $\chi(\bm{Q},\it{E})$ of MBT using an $\it{ab~initio}$ method (Details in SM). To compare with experiments, we also obtain the pairwise exchange parameters $J_{ij}$ from the inverse of susceptibility matrix, $[\chi(\bm Q,E=0)]^{-1}$, with a subsequent Fourier transform~\cite{szczech1995prl, katsnelson2004jpcm, kotani2008jpcm, ke2013prb, ke2020arxiv}. Figure~\ref{fig:DFT} shows the imaginary part of the full dynamic transverse susceptibility, $\mathrm{Im}[\chi(\bf{Q},\it{E})]$, from linear-response conforms to expectations of an insulating local moment system.  The linewidths are sharp since the insulating gap eliminates the low-energy electron-hole (Stoner) excitations responsible for Landau damping of the spin waves.  The calculations show reasonable qualitative agreement with the energy scales of experimental data and numerical extraction of the \textit{ab initio} Heisenberg exchange values, as shown in Fig.~\ref{fig:DFT}(d), supporting the experimental picture of quasi-2D spin dynamics with long-range and competing FM and AF intralayer exchange interactions. DFT informs us that the strong FM $J_1$ arises from Mn-Te-Mn superexchange through $\sim 90^{\circ}$ bond angle and the competing AF $J_2$ interaction must also originate from more complex superexchange paths in the insulator. Interestingly, DFT finds long-range intralayer interactions beyond 10~\AA~in insulating MBT, but their values lack quantitative agreement with experimental results, even with regard to the sign of the exchange.

Along $(0,0,L)$, DFT  and INS data find a similar bandwidth of $\sim$ 0.4 meV, but DFT finds no evidence for a dominant AF NN exchange $J_c$. Rather, the net AF interlayer coupling from DFT is comprised of interactions of many interlayer neighbors, $J_{\text{c}}^{\text{eff}}=1/Z_0\sum_{j, \perp} J^{0j}_{\text{c}} \cdot Z_j$, where $Z_j$ is the number of the $j$-th nearest neighbor, and $Z_0=6$. However the small exchange energies of MBT (often with $|SJ_{ij}| \sim $ 0.01 meV) present a challenge for the numerical accuracy of DFT.

We now discuss the implications of our study of the spin dynamics in MBT. Both INS and DFT directly justify a picture of weakly coupled quasi-two-dimensional FM vdW layers. For the interlayer coupling, while INS can only reasonably fit a single effective interlayer $J_\text{c}$, DFT suggests that this might be an oversimplification and magnetic coupling across the vdW gap may have contributions from multiple neighbor shells. INS measurements on Bi$_{1.99}$Mn$_{0.01}$Te$_{3}$ and magnetization data confirm contributions to the magnetic anisotropy from both single-ion and interlayer two-ion terms. We estimate a $20\%$ reduction of the spin gap at the surface due to the lack of neighbors from surface termination \cite{SM}, which more easily enables the pinning of surface spin-flops states in applied fields as observed in ref.~\cite{Sass2020}. Enhanced surface spin fluctuations will further reduce $T_{\text{N}}$ ($\approx 10\%$) at the surface which is a consideration for any surface spectroscopy measurements. It has recently been proposed that the MBT series, including MnBi$_4$Te$_7$ and MnBi$_6$Te$_{10}$ vdW magnets with non-magnetic spacer layers, possess unique magnetization dynamics controlled by both the magnetic anisotropy and interlayer coupling. Detailed and systematic INS studies are imperative towards understandings of the magnetism in the MBT series.

Finally, observation of long-range interactions and anomalous damping potentially originates from novel coupling of spin waves to topological (chiral) carriers. However, we have demonstrated that such behavior is not inconsistent with insulating MBT containing magnetic vacancy defects. Further study of novel coupling to chiral carriers could be tested by comparing the spin wave dispersion for different carrier densities, such as the charge neutral Mn(Bi$_{1-x}$Sb$_{x}$)$_{2}$Te$_{4}$. However, accounting for any additional chemical and magnetic disorder would have to be considered in such an approach.

\section*{Acknowledgements} \label{sec:acknowledgements}
We are grateful for helpful discussions with Andreas Kreyssig and Peter. P. Orth. This work is supported by the U.S. Department of Energy, Office of Basic Energy Sciences, Division of Materials Sciences and Engineering. Ames Laboratory is operated for the U.S. Department of Energy by Iowa State University under Contract No. DE-AC02-07CH11358. This research used resources at the Spallation Neutron Source, a DOE Office of Science User Facility operated by the Oak Ridge National Laboratory.

\end{document}